\documentstyle[epsf]{mn}

\newif\ifAMStwofonts
\ifoldfss
  \ifCUPmtlplainloaded \else
    \NewTextAlphabet{textbfit} {cmbxti10} {}
    \NewTextAlphabet{textbfss} {cmssbx10} {}
    \NewMathAlphabet{mathbfit} {cmbxti10} {} % for math mode
    \NewMathAlphabet{mathbfss} {cmssbx10} {} %  "   "    "
  \fi
  \ifAMStwofonts
    \ifCUPmtlplainloaded \else
      \NewSymbolFont{upmath} {eurm10}
      \NewSymbolFont{AMSa} {msam10}
      \NewMathSymbol{\upi}     {0}{upmath}{19}
      \NewMathSymbol{\umu}     {0}{upmath}{16}
      \NewMathSymbol{\upartial}{0}{upmath}{40}
      \NewMathSymbol{\leqslant}{3}{AMSa}{36}
      \NewMathSymbol{\geqslant}{3}{AMSa}{3E}

      \let\geq=\geqslant 
    \fi
  \fi
\fi % End of OFSS

\ifnfssone
  \newmathalphabet{\mathit}
  \addtoversion{normal}{\mathit}{cmr}{m}{it}
  \addtoversion{bold}{\mathit}{cmr}{bx}{it}
  \newmathalphabet{\mathbfit} % math mode version of \textbfit{..}
  \addtoversion{normal}{\mathbfit}{cmr}{bx}{it}
  \addtoversion{bold}{\mathbfit}{cmr}{bx}{it}
  \newmathalphabet{\mathbfss} % math mode version of \textbfss{..}
  \addtoversion{normal}{\mathbfss}{cmss}{bx}{n}
  \addtoversion{bold}{\mathbfss}{cmss}{bx}{n}
  \ifAMStwofonts
    \ifCUPmtlplainloaded \else
      \UseAMStwoboldmath
      \makeatletter
      \new@mathgroup\upmath@group
      \define@mathgroup\mv@normal\upmath@group{eur}{m}{n}
      \define@mathgroup\mv@bold\upmath@group{eur}{b}{n}
      \edef\UPM{\hexnumber\upmath@group}
      \new@mathgroup\amsa@group
      \define@mathgroup\mv@normal\amsa@group{msa}{m}{n}
      \define@mathgroup\mv@bold\amsa@group{msa}{m}{n}
      \edef\AMSa{\hexnumber\amsa@group}
      \makeatother
      \mathchardef\upi="0\UPM19
      \mathchardef\umu="0\UPM16
      \mathchardef\upartial="0\UPM40
      \mathchardef\leqslant="3\AMSa36
      \mathchardef\geqslant="3\AMSa3E

      \let\geq=\geqslant 
    \fi
  \fi
\fi % End of NFSS release 1

\ifnfsstwo
  \DeclareMathAlphabet{\mathbfit}{OT1}{cmr}{bx}{it}
  \SetMathAlphabet\mathbfit{bold}{OT1}{cmr}{bx}{it}
  \DeclareMathAlphabet{\mathbfss}{OT1}{cmss}{bx}{n}
  \SetMathAlphabet\mathbfss{bold}{OT1}{cmss}{bx}{n}
  \ifAMStwofonts
    \ifCUPmtlplainloaded \else
      \DeclareSymbolFont{UPM}{U}{eur}{m}{n}
      \SetSymbolFont{UPM}{bold}{U}{eur}{b}{n}
      \DeclareSymbolFont{AMSa}{U}{msa}{m}{n}
      \DeclareMathSymbol{\upi}{0}{UPM}{"19}
      \DeclareMathSymbol{\umu}{0}{UPM}{"16}
      \DeclareMathSymbol{\upartial}{0}{UPM}{"40}
      \DeclareMathSymbol{\leqslant}{3}{AMSa}{"36}
      \DeclareMathSymbol{\geqslant}{3}{AMSa}{"3E}

      \let\geq=\geqslant 
    \fi
  \fi
\fi % End of NFSS release 2

\ifCUPmtlplainloaded \else
  \ifAMStwofonts \else % If no AMS fonts
    \def\upi{\pi}
    \def\umu{\mu}
    \def\upartial{\partial}
  \fi
\fi

\title{Deep IR and optical studies of the fields of 3 proposed
T$\dot{\rm Z}$O remnants}
\author[M.J.Coe and S.L.Pightling]
       {M.J.Coe and S.L.Pightling \\
        Department of Physics and Astronomy, The University,
Southampton, SO17 1BJ, UK.}
\date{Accepted .
      Received ;
      }

\pagerange{\pageref{firstpage}--\pageref{lastpage}}
\pubyear{1997}

\begin{document}

\maketitle

\label{firstpage}

\begin{abstract}

Deep infrared and optical images are presented of three proposed
remnants of Thorne-$\dot{\rm Z}$ytkow Objects (T$\dot{\rm Z}$O). In
particular, the IR data go several infrared magnitudes deeper than
previous observations and in at least one case reveal the existence of
weak objects within the error circles. It is argued, however, that
none of the objects is likely to be the binary companion to the X-ray
source in that region. These data present severe limits on any
possible star or residual envelope at the distances of the respective
X-ray objects.

\end{abstract}


\section{Introduction}

Recent work by several authors, especially van Paradijs et al. (1995)
and Mereghetti \& Stella (1995), have identified a group of pulsating
X-ray sources that appear to have no detectable optical
counterpart. In addition, these objects show no evidence of any
binarity and all have pulse periods in the 5-10s range. Their X-ray
luminosities are all sufficiently high that accretion must be
occurring on to the neutron star from some source other than the
interstellar medium. One possible scenario is that they are the first
objects identifiable as the remains of Thorne-$\dot{\rm Z}$ytkow Objects
(T$\dot{\rm Z}$Os). 

The progenitor T$\dot{\rm Z}$Os would have consisted of a neutron star
embedded in the core of another star (Thorne \&$\dot{\rm Z}$ytkow
1977).  It is thought that these T$\dot{\rm Z}$Os could be formed as a
result of the evolution of high mass systems through a common envelope
phase (Ghosh et al. 1997). Once formed, thermonuclear burning via the
{\it r-p} process occurs in the core region of the T$\dot{\rm
Z}$O. This supports the highly convective outer envelope against
gravitational forces.  However, once there is insufficient fuel
remaining for the {\it r-p} process to occur efficiently, the outer
envelope collapses on the Kelvin-Helmholtz timescale (Podsiadlowski,
Cannon and Rees, 1995). Some of the envelope material then forms into
a disk around the neutron star with a radius that is dependent on the
amount of angular momentum in the system. It is from this disk of
material that the neutron star may accrete enough matter to power the
X-ray pulsations seen.

In order to explore this possibility deep infrared imaging
observations have been carried out from UKIRT and the NASA IRTF to
search for any IR emission from such a disk around three of the neutron
stars. The observations reach as deep as J$\sim$20 and in some cases reveal
the presence of faint objects in the X-ray error circles. However, the
characteristics of the IR colours, and other considerations, strongly
suggest that these are not the TZO envelopes around the neutron
stars. However, the new stringent limits
on any IR companion in these systems leave these objects as enigmatic as
ever.

The three systems investigated here are:

\subsection{1E2259+586} 

The X-ray pulsar 1E2259+586 lies in the centre of the supernova remnat
G109.1-1.0. The most recent estimates on its age make it approximately
3000 years old (Parmar {\it et al.} 1997), though values up to 10000
years have been quoted. G109.1-1.0 is at an estimated distance of
about 4 kpc. Fahlman{\it et al.} (1982) demonstrated that the X-ray
column densities to both the pulsar and the SNR were similar and hence
that these two objects were probably related.  However, the latest
results from Parmar et al.(1997) indicate that this may not be the
case. They find the X-ray column density to the pulsar to be greater
than that to the supernova remnant. They suggest that this could be
due to the pulsar being at a greater distance than the supernova
remnant or to the presence of absorbing material local to the pulsar.

However, assuming the same distance to 1E2259+586 as the SNR implies a 
pulsar X-ray luminosity of $\approx10^{28}$ Watts.  Results from {\it EXOSAT},
{\it GINGA} and {\it BeppoSAX} (Hanson et al. 1988, Koyama et al. 1989 and
Parmar et al. 1997 respectively) have shown that the 7s period of 1E2259+586 is
slowly increasing, but the spindown is too slow to power the observed
luminosity.  There is no evidence of binary motion, the upper limit on
$a_{x} \sin i$ from pulse timing measurements being as low as 30
light-ms (Mereghetti, Israel and Stella, 1998).  Early searches for the optical
counterpart (Fahlman {\it et al.} 1982; Margon \& Anderson 1983) found
several possible candidates, one of which (star D) was tentatively
identified with 1E2259+586 by Middleditch, Pennypacker \& Burns (1983)
on the basis of IR pulsations.  However, more sensitive fast
photometric observations by Davies {\it et al.} (1989) failed to find
any pulsations, and there would now appear to be little positive
evidence associating star D with 1E2259+586.  A fainter, multicolour
photometric study of all the possible counterparts by Davies \& Coe
(1991), together with revised astrometry, updated the list of
possible counterparts.
 
Coe \& Jones (1992) presented the first optical spectroscopy of
all the candidates brighter than V=23 and re-analysed the Einstein X-ray
images of this source taken in 1981 in order to check the position and size
of the X-ray error circle. They found that the relocation and reduction in
the size of the error circle stemming from the re-processing of these data
considerably changes the perspective on possible counterparts to the X-ray
pulsar. In addition, a careful study of the images reveals previously
unreported features that are possibly associated with jet activity from the
pulsar, a possibility originally suggested by Gregory and Fahlman
(1980). However, Hurford \&
Fesen 1995 report subsequent {\it Rosat} HRI observations which, while
confirming the existence of such features, they suggest are
just statistical fluctuations in the SNR emission.

\subsection{4U0142+62}

This persistent source has appeared in most X-ray catalogues and was
originally confused (due to poor angular resolution) with the Rosat
source RX J0146.9+6121 (=LSI +61 235) - the two sources are only 24
arcminutes apart. Measurements by White et al. (1987) with {\it
EXOSAT} reported a 25m modulation from the region (later shown to be
coming from LSI +61 235 (Mereghetti, Stella \& De Nile 1993)) but
using the unmodulated lower energy data they were able to identify the
position of 4U0142+62. The detection of X-ray pulsations at 8.7s was
reported by Israel et al. (1994) who concluded that they were most
likely coming from the optically unidentified system 4U0142+62 rather
than the new {\it Rosat} source. Further measurements by White et
al. (1996) using {\it ASCA} confirmed this conclusion and also
demonstrated evidence for an excess in the X-ray halo around the
source. This could be due to material local to the pulsar along the
line of site, possibly in the form of a molecular cloud or perhaps the
remains of a TZO envelope. No evidence of any binary motion has been
reported from the X-ray signal.

The accurate position reported by White et al. (1987) allowed them to
locate the X-ray error circles from both {\it EXOSAT} and {\it
Einstein} on the sky. As in the case of 1E2259+586 there is no
evidence of any obvious optical candidates within the defined
regions. They obtained an R band image and set a limit of R$\geq$22.5
for any counterpart. The ASCA observations (White et al. 1996)
determined the column to the source to be 8 x 10${^{21}}$ cm${^{-2}}$
which in turn allows them to deduce an optical extinction of
A$_{v}$$\sim$4.7. The distance to the source is not well defined but
probably lies in the range 0.5--2.0 kpc. As in the case of 1E2259+586
a molecular cloud lies near to, or in front of 4U0142+62 (Mereghetti
\& Stella 1995) and may be affecting the column measurement.

\subsection{RX J1838.4--0301}

This source was originally reported by Schwentker (1994) as 5.5s
pulsar possibly associated with a SNR at a distance of
4kpc. Interestingly, in this case the X-ray positional error circle
was reported to include a $\sim$14 mag optical candidate. Subsequently
Mereghetti et al. (1997) reanalysed the same {\it Rosat} data and
reported are different interpretation of the data. They obtained an
optical spectrum of the star in the error circle and concluded that
this object was simply a K5 main sequence star and that the X-ray
emission originated from its corona .

\section{The observations}

\subsection{1E2259+586}

This source was observed by UKIRT using IRCAM3 on 1996 October 23 in
Service mode. The plate scale was $\sim$0.3 arcsec per pixel giving an
image size of 73 x 73 arcsec. Observations were taken in both J and K
bands using a standard mosaicing technique with a total exposure of 27
minutes in each band. The K band image of the field is shown in Figure
1.

\begin{figure}
\begin{center}
{\epsfxsize 0.99\hsize
 \leavevmode
 \epsffile{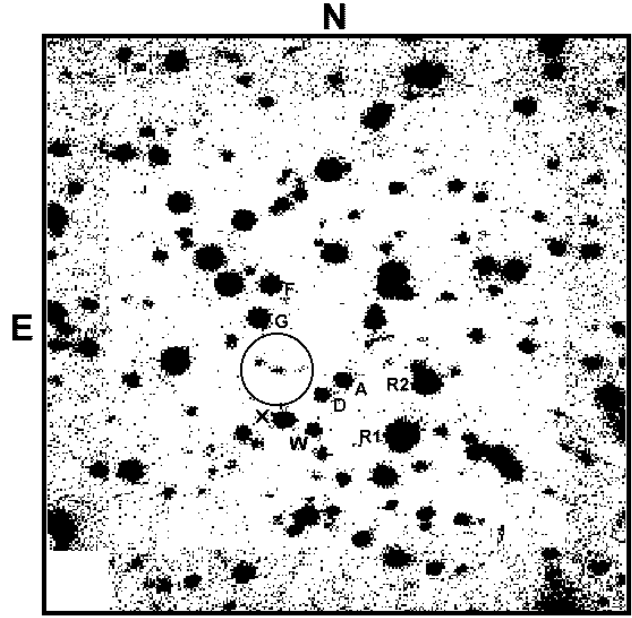 }
}\end{center}
%\vspace{11cm}
\caption{A K band image of the field of 1E2259+586. The field size is
73 arcseconds and the circle shown represents the {\it Rosat} X-ray
error circle of radius 5 arcsec (Dennerl, private communication). The
three very faint objects within this error circle are referred to in
the text as objects L, K and N from East to West respectively.}
\label{}
\end{figure}

Superimposed on the image is the {\it Rosat} error circle (Dennerl,
private communication) and the labels for several of the brighter
stars in the region (following the convention of Coe et al. 1994a). The
image is approximately 1-2 magnitudes deeper than any previous
measurements and for the first time permits measurements of the three
faint objects L, K and N.

In order to obtain an accurate calibration of the images, the
established objects R1, R2, G, F, A and D (all shown in the figure)
were used. The instrument magnitudes of these objects were compared
with the reported fluxes in Davies \& Coe (1991) and an excellent
correlation was obtained (this, incidentally, confirms that there has
been no detectable variation in the fluxes from any of these
objects). Using this correlation the magnitudes for the three
candidates L, K and N, as well as the nearby objects X and W, were
determined. The results are presented in Table 1. From sampling the
background around the area of interest the three sigma upper limits for
any undetected source in this field may be determined to be
J$\geq$19.6 and K$\geq$18.4.

\begin{table}
\caption{Infrared observational measurements of the candidates for 1E2259+586}
\begin{tabular}{cccc}

Object & J band & K band & (J-K) \\
&&&\\
X & 17.0$\pm$0.1 & 16.1$\pm$0.06 & 0.89$\pm$0.12 \\
W & 18.1$\pm$0.1 & 17.0$\pm$0.1 & 1.1$\pm$0.14  \\
L & 20.1$\pm$0.5 & 17.9$\pm$0.4 & 2.2$\pm$0.6  \\
K & 20.0$\pm$0.5 & 18.0$\pm$0.4 & 2.0$\pm$0.6  \\
N & $\geq$20.6   & 18.3$\pm$0.4 & $\geq$2.3 \\
\end{tabular}
\end{table}

\subsection{4U0142+62}

This source was observed by the NASA IRTF on 1996 August 25 using the
NSFCAM imager. This instrument has a plate scale of 0.3 arcsec per
pixel giving an image size of 77 arcsec. Observations were taken in
the J and K bands of 30 minutes each. In addition to the target field
the standard star HD18881 was observed together with an appropiate number
of flats.

The K band image of the field is shown in Figure 2. Also shown in this
figure is the {\it Rosat} positional error circle from Hellier
(1994). As one can see the field is extremely empty lacking any
possible candidates close to the X-ray position. The faintest object
detected and measured had a magnitude of J=19.6$\pm$0.1 and K=16.88$\pm$0.07.
The measurements of all the objects identified in the figure are given
in Table 2 (note that star identification scheme is an extension of
that started by White et al 1987).

\begin{figure}
\begin{center}
{\leavevmode
 \epsfxsize 0.99\hsize
 \epsffile{ 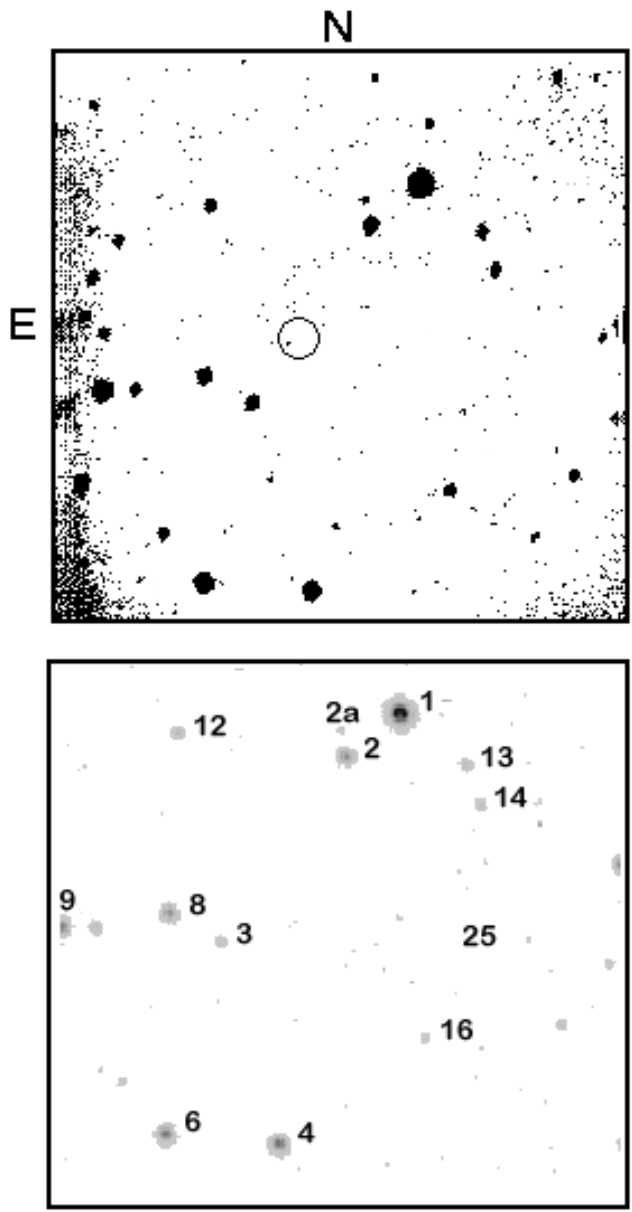 }
}\end{center}
%\vspace{11cm}
\caption{The upper panel shows a K band image of the field of 4U0142+62. The field size
illustrated is
71 arcseconds and the circle shown represents the {\it Rosat} X-ray
error circle of radius 3.2 arcsec (White et al. 1987). The lower panel
is a V band image (Steinle, private communication) indicating the
identification of stars in the region of the X-ray error circle.}
\label{}
\end{figure}

\begin{table}
\caption{Infrared observational measurements of the candidates for 4U0142+62}
\begin{tabular}{cccc}

Object & J band & K band & (J-K) \\
&&&\\
1 & 13.501$\pm$0.005 & 13.126$\pm$0.004 & 0.375$\pm$0.006 \\
2 & 15.33$\pm$0.01   & 14.91 $\pm$0.01  & 0.418$\pm$0.015  \\
2a & 17.61$\pm$0.04  & 17.09 $\pm$0.03  & 0.513$\pm$0.05  \\
3 & 15.92 $\pm$0.02  & 15.16 $\pm$0.01  & 0.76 $\pm$0.02  \\
4 & 15.14$\pm$0.01   & 14.86 $\pm$0.01  & 0.27 $\pm$0.01  \\
6 & 14.95$\pm$0.01   & 14.63 $\pm$0.01  & 0.32 $\pm$0.01  \\
8 & 15.74$\pm$0.01   & 15.44 $\pm$0.01  & 0.30 $\pm$0.02  \\
9 & 15.52$\pm$0.01   & 15.20 $\pm$0.01  & 0.32 $\pm$0.02  \\
13 & 16.82$\pm$0.02  & 16.38 $\pm$0.02  & 0.43 $\pm$0.03  \\
14 & 17.15$\pm$0.03  & 16.67 $\pm$0.03  & 0.47 $\pm$0.04  \\
16 & 17.02$\pm$0.03  & 16.38 $\pm$0.02  & 0.64 $\pm$0.03  \\
25 & 18.55$\pm$0.06  & 17.54 $\pm$0.04  & 1.01 $\pm$0.07  \\
29 & 17.05$\pm$0.03  & 16.65 $\pm$0.02  & 0.40 $\pm$0.04  \\
30 & 17.16$\pm$0.03  & 16.34 $\pm$0.02  & 0.82 $\pm$0.03  \\
\end{tabular}
\end{table}

\subsection{RX J1838.4-0301}

Notwithstanding the discussion in Section 1 of this paper, this source
lacks a securely identified counterpart. Even the question of
the existence, or otherwise, of X-ray pulsations needs resolving. It was
therefore felt appropiate to explore the region around the error
circle to see if any sources revealed themself as typical of a
counterpart to an X-ray source. Consequently, on 1996 November 15
images were taken in each of the B,V,R and I band using the Tek8 CCD
camera on the 1m SAAO telescope with standard Johnson filters. In
addition an image was recorded through the H$\alpha$ filter. The
exposure times for B,V and I were 300s, while R was 200s and H$\alpha$
was 900s.

If the X-ray pulsations are real then one obvious explanation for this
system is that it is either a Be/X-ray binary system or a supergiant
system. If this is the case then the optical counterpart would be a
strong source of H$\alpha$ emision. To investigate this possibility
all the objects on the frame were measured in both the R band and
H$\alpha$ and the two sets of instrument magnitudes plotted against
each other. This approach has been very successful in identifying
counterparts to high mass X-ray binaries (see, for example, Coe et
al., 1994b).  In this case the result revealed no obvious source of
H$\alpha$ emission in the field.

In addition to the above, deep IR observations were obtained from
UKIRT using IRCAM3 on 1997 September 13 (J band) and 1997 September 29
(K band). Total exposures of 18 minutes were used in both bands. The
instrument set up was identical to the observations of 1E2259+586. The
K band image is shown in Figure 3 along with the determinations of the
position of the {\it Rosat} source by Schwentker (1994) and Mereghetti
et al. (private communication and 1997). The J and K magnitudes were
measured of all the identified objects shown in Figure 3 and the
results are presented in Table 3.

\begin{figure}
\begin{center}
{\epsfxsize 0.95\hsize
 \leavevmode
 \epsffile{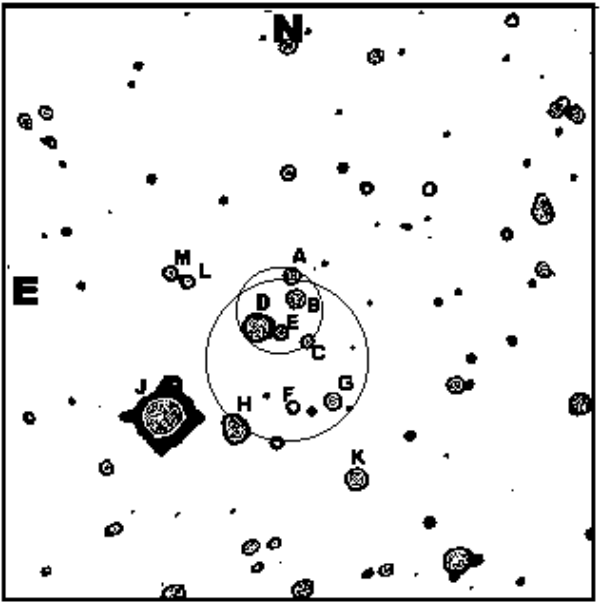 }
}\end{center}
%\vspace{11cm}
\caption{K band image of RX J1838.4-0301. The field size is 73
arcseconds and the small circle of radius 5'' shows the estimate of
the location of the {\it Rosat} source (Mereghetti et al., private
communication and 1997), whereas the larger circle (radius 10'') is that of
Schwentker (1994).}
\label{}
\end{figure}

\begin{table}
\caption{Infrared observational measurements of the candidates for RX J1838.4-0301}
\begin{tabular}{cccc}

Object & J band & K band & (J-K) \\
&&&\\
A & 12.69$\pm$0.03 & 12.74$\pm$0.03 & -0.05$\pm$0.04 \\
B & 12.29$\pm$0.03 & 12.62$\pm$0.03 & -0.33$\pm$0.04  \\
C & 14.01$\pm$0.03 & 14.69$\pm$0.03 & -0.68$\pm$0.04  \\
D & 12.03$\pm$0.03 & 11.98$\pm$0.03 & +0.05$\pm$0.04  \\
E & 13.72$\pm$0.03 & 14.16$\pm$0.03 & -0.44$\pm$0.04  \\
F & 13.38$\pm$0.03 & 13.63$\pm$0.03 & -0.25$\pm$0.04  \\
G & 13.36$\pm$0.03 & 13.24$\pm$0.03 & +0.12$\pm$0.04  \\
H & 12.17$\pm$0.03 & 12.24$\pm$0.03 & -0.07$\pm$0.04  \\
J & 11.55$\pm$0.03 & 12.01$\pm$0.03 & -0.46$\pm$0.04  \\
K & 12.47$\pm$0.03 & 12.19$\pm$0.03 & +0.28$\pm$0.04  \\
L & 13.19$\pm$0.03 & 14.15$\pm$0.03 & -0.96$\pm$0.04  \\
M & 13.61$\pm$0.03 & 14.24$\pm$0.03 & -0.63$\pm$0.04  \\
\end{tabular}
\end{table}

It is obvious from the UKIRT image that the reported ``14th magnitude
star'' is in fact a complex of 5 or more previously unresolved
stars. The limits on any undetected sources in the field were
determined by sampling the background around the X-ray error
circle. The resulting three sigma upper limits are J$\geq$17.7 and
K$\geq$16.9.

\section{Discussion}

In order to investigate the nature of the objects in the fields of
our targets an infrared colour plot was produced for each
region. Figure 4 shows the plot for the first source, 1E2259+586.

\begin{figure}
\begin{center}
{\epsfxsize 0.99\hsize
 \leavevmode
 \epsffile{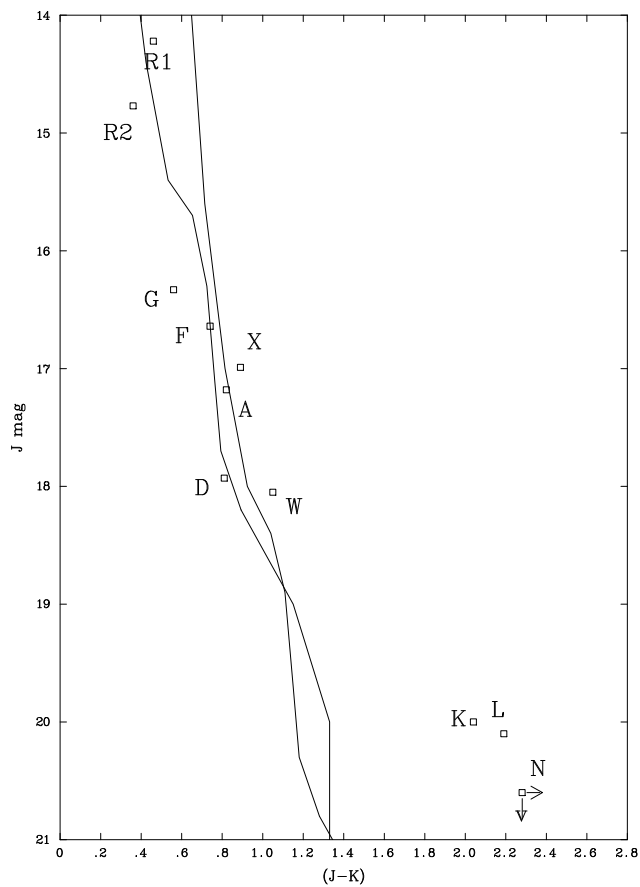} 
}\end{center}
%\vspace{11cm}
\caption{The infrared colours (J--K) for all the objects close to, or
in, the X-ray error circle of 1E2259+586. The solid and dotted lines
represent the location of the Main Sequence at reddenings
of A$_{v}$=2.5 and 4.7 respectively.}
\label{}
\end{figure}

To help interpret this diagram the location of the Main Sequence (MS) has
been determined with the appropiate reddening. The solid line in the
figure shows the MS with a reddening of A$_{v}$=2.5 (corresponding to a
distance of 4kpc), while the dotted line is for A$_{v}$=4.7 (corresponding
to 10kpc). These two values for reddening encompass the range of
distances proposed for 1E2259+586, though the smaller value is the more
probable (Davies \& Coe 1991). As one can see, the reddening
approximately follows the direction of the MS and hence does not
provide a very useful distance discriminator. However, it is clear
that most of the more obvious sources lie comfortably along the MS
line and present no unusual IR colours. Object E, discussed in Coe \&
Jones (1992) as a possible white dwarf, is not detected in either IR
image and hence is fainter than J=21 and K=18.5.

The same cannot be said though for the new objects labelled K, L and N
- these are the objects which actually lie in the {\it Rosat} error
circle. A very large amount of reddening is required to make these
objects consistent with a MS object - for example, a cluster of young
B0 stars would need A$_{v}$$\sim$13--14. Though it is highly
improbable that the reddening in this galactic direction could, in
general, be larger than A$_{v}$=$\sim$5, such a high value could be
consistent with a large amount of local material around a distant
cluster.

Alternatively, A$_{v}$=4.7 corresponds to E(J--K)=0.8 (Rieke \&
Lebofsky 1985) and the observed colour would be consistent with a
Seyfert galaxy which typically has (J--K)$_{o}$=1.6, and hence would
be reddenned to (J--K)=2.4. However, the chance of three such objects
lying within 10 arcsec seems unlikely.

For the field containing 4U0142+61 the brightest 14 objects have had
both their J and K colours determined. The results are presented in
Figure 5.

\begin{figure}
\begin{center}
{\epsfxsize 0.99\hsize
 \leavevmode
 \epsffile{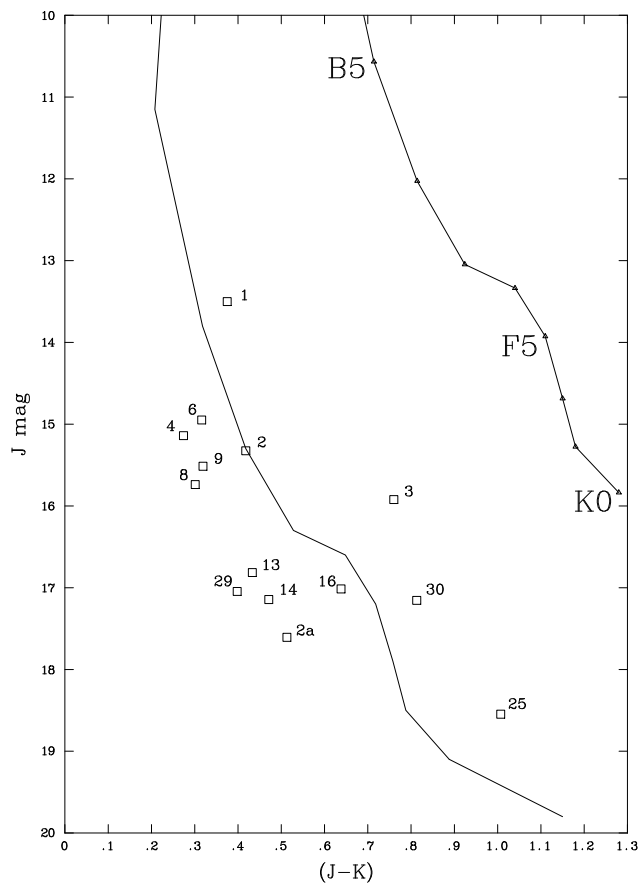}
}\end{center}
%\vspace{11cm}
\caption{The infrared colours (J--K) for all the objects close to 
the X-ray error circle of 4U0142+62. The solid line
represent the location of the Main Sequence at a reddening of
A$_{v}$=4.7 and a distance of 1kpc, and the dotted line the MS with
A$_{v}$=2.4 at distance of 6kpc.}
\label{}
\end{figure}

Various fits to the data were undertaken to obtain the best estimate
of the distance and reddening to the field stars. The solid line in
the figure shows the best estimate for the Main Sequence at the
distance (1kpc) and reddening (A$_{v}$=4.7) quoted in White et al
(1996). Clearly the field stars do not lie anywhere close to this
position, nor does simply reducing the reddening to, for example,
A$_{v}$=2 (analagous perhaps to a position just in front of a molecular
cloud reported by Mereghetti \& Stella 1995). The best fit was
obtained for an A$_{v}$=2.4 at distance of 6kpc - indicated in Figure
5 by the dashed line. From this one can conclude that none of the
visible stars are related to the X-ray source - perhaps not too
surprising since they all lie well away from the X-ray error circle.

In the case of RX J1838.4-0301 the lack of any obvious H$\alpha$
source in the immediate region of the X-ray position could be
interpretted as support for the conclusions of Mereghetti et
al. (1997) that the X-ray signal is coronal in nature. However, the IR
images reported here clearly show the presence of several comparably
bright stars in the region of the {\it Rosat} source. Figure 6 shows a
similar plot to Figures 4 and 5, but for the objects in the field of RX
J1838.4-0301.

\begin{figure}
\begin{center}
{\epsfxsize 0.99\hsize
 \leavevmode
 \epsffile{ 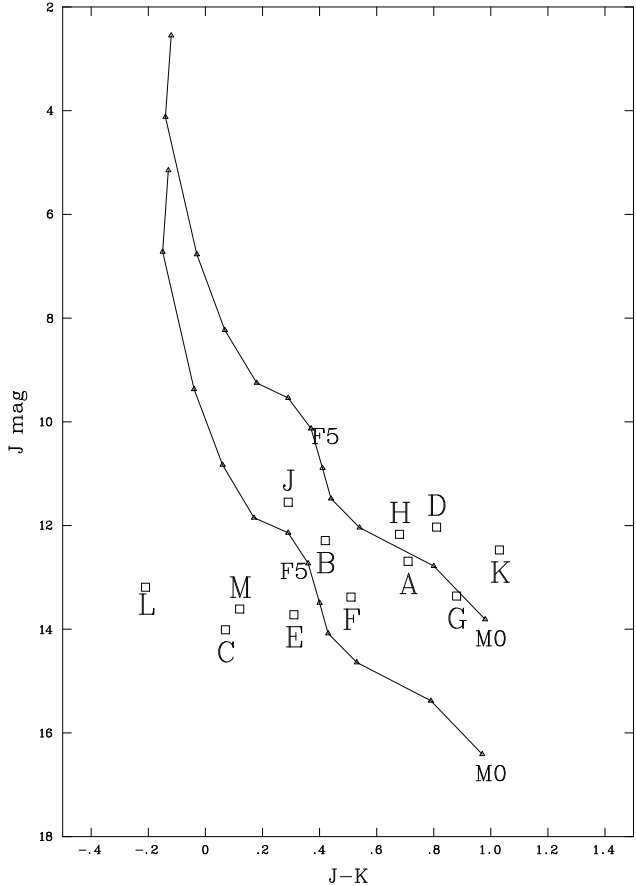 }
}\end{center}
%\vspace{11cm}
\caption{The infrared colours (J--K) for all the objects close to, or
in, the X-ray error circle of RX J1838.4-0301. The solid line
represent the location of the Main Sequence at a reddening of
A$_{v}$=0.42 and a distance of 0.3kpc. The dashed line is the MS at
the same A$_{v}$, but at a distance of 1.0kpc.}
\label{}
\end{figure}

Schwentker (1994) reports a hydrogen column to this source in the
range (1-3) x 10$^{21}$ atoms per sq.cm. If we use this value with the
relationship given in Ryter, Cesarsky \& Audouze (1975) we can deduce
an E(B-V) in the range 0.2 - 0.4 and hence an E(J-K) in the range 0.1
- 0.2. The locus of the Main Sequence at an average reddening of
E(J-K)=0.15 (i.e. E(B-V)=0.27) and the distance of 1kpc suggested by
Schwentker (1994) is marked by the dashed line in Figure 6. Many of the
stars in the region of the X-ray source lie close to this line, but
the spread is quite significant in the (J-K) axes. Mereghetti et al
(1997) have proposed a K5 star at 300pc as being consistent with the
X-ray data (if the pulsations are not real). A study of all the stars
in this direction indicates that at distances beyond 200-300 pc the
hydrogen column will always be of the order 10$^{21}$ atoms per
sq.cm. The solid line in Figure 6 shows the reddening expected for
such a column, but now at a distance of 300pc. Star A lies extremely
close to the position expected for a K5 star at such a distance and
reddening, and this object may have been the dominating star in
Mereghetti et al's data. However, the other stars in the {\it Rosat}
error circle lie scattered all over the diagram and offer many other
possible solutions. For examply, Star D is the brightest in our IR
data at J=12.0 and probably lies even closer than Star A. Certainly
Mereghetti et al might well have had several of the stars A-E in the
2'' slit for their spectroscopy.

If the pulsations turn out to be real as proposed by Schwentker
(1994), then it would support RX J1838.4-0301 remaining a member of
this class of enigmatic objects. What is clear is that it is very
unlikely to be a member of any Be or supergiant binary system. Further
X-ray observations are clearly required both to refine the X-ray
position in this crowded field, and to clarify the situation regarding
the possible X-ray pulsations.

Finally, a rough estimate of the limiting size for a disk producing
black body emission may be obtained from the IR results presented
here. In the case of 1E2259+586 the limits of J$\geq$20 and
K$\geq$18.5 imply a limiting size of approximately 1 solar radius (7 x
$10^{8}$m) for a 2500K black body at 4.5kpc. Podsiadlowski, Cannon and
Rees (1995) quote a size of the order 3 x $10^{8}$m for a system that
had an initial binary orbital period of 10d and a companion of 15
solar masses. The IR limits for 4U0142+62 are similar and imply a
somewhat smaller disk limit if the distance to the source is at the
lower end of the proposed range of (i.e. 0.5 kpc) - maybe by up to a
factor of 10 (i.e. a disk size of 7 x $10^{7}$m). However, the upper
bounds to the distance (1.5 kpc) give limits similar to the disk sizes
of Podsiadlowski, Cannon and Rees (1995).  Thus these new IR values
are probing the expected disk sizes for a proposed T$\dot{\rm Z}$O
envelope.

\section{Conclusions}

The first deep infrared and optical images have been presented of
three proposed remnants of Thorne-$\dot{\rm Z}$ytkow Objects
(T$\dot{\rm Z}$O). In particular, the IR data go several infrared
magnitudes deeper than previous observations and in at least one case,
1E2259+586, reveal the existence of weak objects within the X-ray
error circle. It is argued, however, that in that case none of the
objects seen is likely to be the binary
companion to the X-ray source in that region. These data present new
limits on any possible IR counterpart to the respective X-ray objects.

\section*{Acknowledgments}

We are very grateful to UKIRT for obtaining data on two of these
sources through their Service programme, and also to the NASA IRTF for
the data on 4U0142+62 obtained in a similar manner. We thank Dave
Buckley for obtaining the optical CCD frames of RX J1838.4-0301 and to
Sandro Mereghetti for many helpful comments. SLP acknowledges the
support of a PPARC studentship.

\bsp

\label{lastpage}

\end{document}